\documentclass[letters,fleqn,usenatbib,useAMS]{mnras}

\usepackage[T1]{fontenc}
\usepackage{ae,aecompl}

\usepackage{graphicx}
\usepackage{color}
\usepackage{xcolor}
\usepackage{amsmath}	
\usepackage{amssymb}	
\usepackage[normalem]{ulem}

\definecolor{ochre}{rgb}{0.8, 0.47, 0.13}

\title[The magnetic field of TIC\,470710327]{A magnetic tertiary in the most massive compact triple-star system}

\stepcounter{footnote}

\author[Hubrig et al.]{
S.~Hubrig$^{1}$,
A.~Vigna-G\'omez$^{2}$,
S.~P.~J\"arvinen$^{1}$,
M.~Sch\"oller$^{3}$,
I.~Ilyin$^{1}$,
\\
$^{1}${Leibniz-Institut f\"ur Astrophysik Potsdam (AIP), An der Sternwarte~16, 14482~Potsdam, Germany} \\
$^{2}${Max-Planck-Institut f\"ur Astrophysik, Karl-Schwarzschild-Str.~1, 85748 Garching bei M\"unchen, Germany}\\
$^{3}${European Southern Observatory, Karl-Schwarzschild-Str.~2, 85748 Garching, Germany}
}

\date{Accepted XXX. Received YYY; in original form ZZZ}

\pubyear{2025}

\begin{document}
\label{firstpage}
\pagerange{\pageref{firstpage}--\pageref{lastpage}}
\maketitle

\begin{abstract}
The system TIC\,470710327
is comprised of three main-sequence OB stars, with an inner compact 1.10\,d eclipsing binary and a
non-eclipsing tertiary on a 52.04\,d orbit. With the tertiary mass of 14.5--16\,$M_{\odot}$ and
both components in the inner eclipsing binary with individual masses of 6--7 and 5.5--6.3\,$M_{\odot}$, it
is currently the most massive compact system known. 
The formation scenario of such a compact triple is uncertain.
It has been suggested that `2 + 2' quadruple dynamics can lead to a stellar merger in the initially more massive binary
and finally result in a highly magnetized tertiary.
Our study confirms the presence of a kG-order magnetic field in the tertiary and the slow
rotation typical for massive magnetic stars. We conclude that finding massive merger candidates by
studies of dynamics in compact, multiple-star systems is an efficient way to understand the
evolution of massive stellar multiplicity and the generation of magnetic fields. 
\end{abstract}




\section{Introduction}
\label{sec:intro}

TIC\,470710327 (BD\,$+$61\,2536) is a massive compact hierarchical triple-star system initially identified
as an early B-type star by \citet{Brodskaya1953}  and as a short period eclipsing binary by \citet{Laur2017},
 both using photometric data. 
More recently, \citet{Eisner2022} used spectra from the HERMES spectrograph combined with eclipse-timing variations
to confirm that TIC\,470710327 consists of three main-sequence OB stars.
The system features an eclipsing binary with a period of 1.10\,d and a non-eclipsing tertiary on a 52.04\,d orbit.
The tertiary star with a mass of 14.5--16\,$M_{\odot}$ is significantly more
massive than both stars in the inner eclipsing binary with individual masses of 6--7 and 5.5--6.3\,$M_{\odot}$. 
Spectral lines belonging to the components of the inner binary are not visible in the HERMES spectra due to their lower masses.
With a mutual inclination of $i=16.8^{+4.2}_{-1.4}\degr$ between the inner and
outer orbit of the system and an eccentricity of $e=0.3$, as reported by  \citet{Eisner2022},
this triple system appears currently dynamically stable. 
The future evolution of TIC\,470710327 is predicted to lead to a mass transfer episode
from the tertiary towards the inner binary,
and while its fate is uncertain,
it could be the progenitor of a binary neutron star or a Thorne-\.Zytkow object \citep{Eisner2022}.
Regarding the potential origin of TIC\,470710327, the authors propose that dedicated radiative hydrodynamical modeling could
determine whether the system, as observed today, formed through core fragmentation, disk fragmentation, or a combination of both.

Although studies of compact hierarchical systems with known period ratios and mutual orbit orientations offer valuable insights
into star formation mechanisms, such systems remain rare. 
However, discoveries are on the rise thanks to the Kepler and Transiting Exoplanet Survey Satellite (TESS) missions
\citep[e.g.,][]{Borkovits2016,Borkovits2019,Borkovits2020,Rappaport2022,Rappaport2023}.
For example, the discovery of the triply eclipsing triple system TIC\,290061484, with the shortest outer period of 24.5 days and
component masses of 6.85, 6.11, and 7.90\,$M_{\odot}$ has recently been reported by \citet{Kostov2024}
using TESS data.
Unfortunately, photometrically discovered compact hierarchical systems are usually very faint and do not allow to obtain
spectral observations and the precise radial velocities necessary to reliably constrain the mutual inclinations and
characterise the physical properties of the components. The reported photometric studies fully rely on a photodynamical
analysis, which is not independent of astrophysical assumptions (e.g.\ \citealt{Rappaport2024}).
Especially mutual inclinations are important to define the degree and character of dynamical interaction between the
inner and outer orbits, but are measured only for a small subset of currently known compact hierarchies owing to the
observational limitations \citep{Tokovinin2021}.

As TIC\,470710327 is relatively bright, it is currently one of the best studied compact hierarchical systems.
Furthermore, to our knowledge, it is the most massive compact triple, with a total system mass of about 23--29\,$M_{\odot}$.
To understand the formation of the system, whose component stars would not fit the current orbit throughout the
pre-main sequence,  \citet{Vigna2022} suggested a progenitor scenario in which `2+2' quadruple dynamics resulted in the
merger of a hypothetical, slightly more massive binary, which created the massive tertiary of TIC\,470710327.
In this proposed scenario, both binary systems form similarly. 
Dynamical interactions trigger the merger of the more massive binary, either during the late stages of star formation or several
million years after the stars reach the zero-age main sequence as they expand on nuclear timescales.

Assuming that the most massive star is a merger product, we highlight several predictions from theoretical studies on massive stellar mergers.
First, the most massive star is magnetic, with a surface magnetic field ranging between 1--10\,kG \citep[e.g.,][]{Schneider2019}.
Second, owing to post-merger angular momentum redistribution \citep{Schneider2019}
and the effects of magnetic wind braking \citep[e.g.,][]{Keszthelyi2021,Wang2022}, the star rotates slowly.
Finally, the stellar surface is predicted to exhibit significant nitrogen enrichment. 
However, not all magnetic stars display a nitrogen excess. Establishing a correlation between the amount of CNO-cycle processed
material at the surface and the field strength of magnetic stars would strongly support the connection between deep mixing and
magnetic phenomena (e.g., Morel 2008). 
To date, such a study has not been conducted.

In this \textit{Letter}, we present our search for the presence of a mean longitudinal magnetic field in TIC\,470710327 using
high-resolution spectropolarimetric observations with the Potsdam Echelle Polarimetric and Spectroscopic Instrument
(PEPSI) at the 2$\times$8.4\,m Large Binocular Telescope (LBT) in Arizona. 
In Section~\ref{sec:obs} we describe our observations, the methodology of our 
analysis, and present the results of the magnetic field measurements.
In Section~\ref{sec:disc} we discuss the implication of our results on future studies of compact
hierarchical triple-star systems.

\section{Observations and measurement results}
\label{sec:obs}

\begin{table*}
\caption{
The logbook of the PEPSI observations and the results of the magnetic field measurements for TIC\,470710327.
The first column gives the Barycentric Coordinate Time of the mid-exposures,
followed by the signal-to-noise ratio measured in the Stokes~$I$ spectra
in the spectral region around 6100\,\AA{} in Column~2.
The line mask used, the radial velocities measured, the false alarm probability (FAP) values, the detection flag
(where DD means definite detection, MD marginal detection, and ND no detection),
the measured mean longitudinal magnetic field strength using the LSD technique, and the
corresponding orbital phase calculated assuming the initial epoch {\bf ${\rm BJD}=2458880.394$} from \citet{Eisner2022}
are presented in Columns~3--8.
}
\label{tab:obsall}
\centering
\begin{tabular}{lcc r@{$\pm$}l cc r@{$\pm$}lc}
\hline
\multicolumn{1}{c}{BJD} &
\multicolumn{1}{c}{$S/N$} &
\multicolumn{1}{c}{Line} &
\multicolumn{2}{c}{RV} &
\multicolumn{1}{c}{FAP}&
\multicolumn{1}{c}{Det.} &
\multicolumn{2}{c}{$\left< B \right>_{\rm z}$} &
\multicolumn{1}{c}{Orbital}\\
\multicolumn{1}{c}{} &
\multicolumn{1}{c}{} &
\multicolumn{1}{c}{mask} &
\multicolumn{2}{c}{(km\,s$^{-1}$)} &
\multicolumn{1}{c}{} &
\multicolumn{1}{c}{flag} &
\multicolumn{2}{c}{(G)}&
\multicolumn{1}{c}{phase}  \\
\hline
2460281.705& 472 & \ion{He}{ii}, \ion{O}{ii}                 & $-$51.62 & 2.25 & $4.8\times10^{-6}$  & DD & 709& 99 & 0.926  \\
2460597.908& 390 & \ion{He}{i}, \ion{C}{iv}, \ion{O}{ii/iii} & $-$44.86 & 1.31 & $9.9\times10^{-6}$  & DD &  12&31  & 0.001\\
2460599.866& 463 & \ion{He}{i/ii}, \ion{O}{ii}, \ion{Si}{iv} & $-$53.79 & 1.65 & $7.2\times10^{-4}$  & MD & 591 & 126 & 0.039\\
\hline
\end{tabular}
\end{table*}

\begin{figure*}
 \centering
\includegraphics[width=0.44\textwidth]{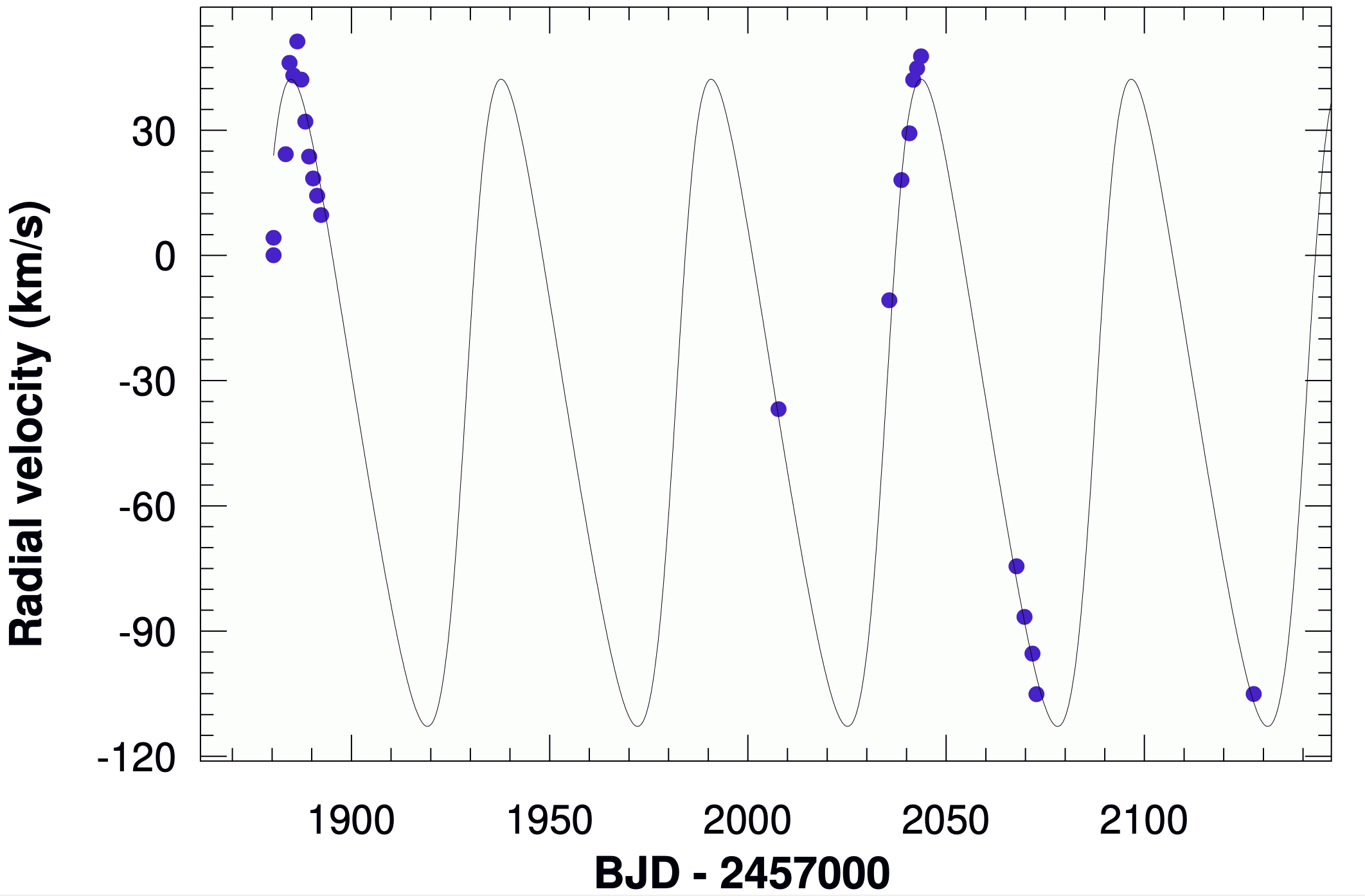}
\includegraphics[width=0.44\textwidth]{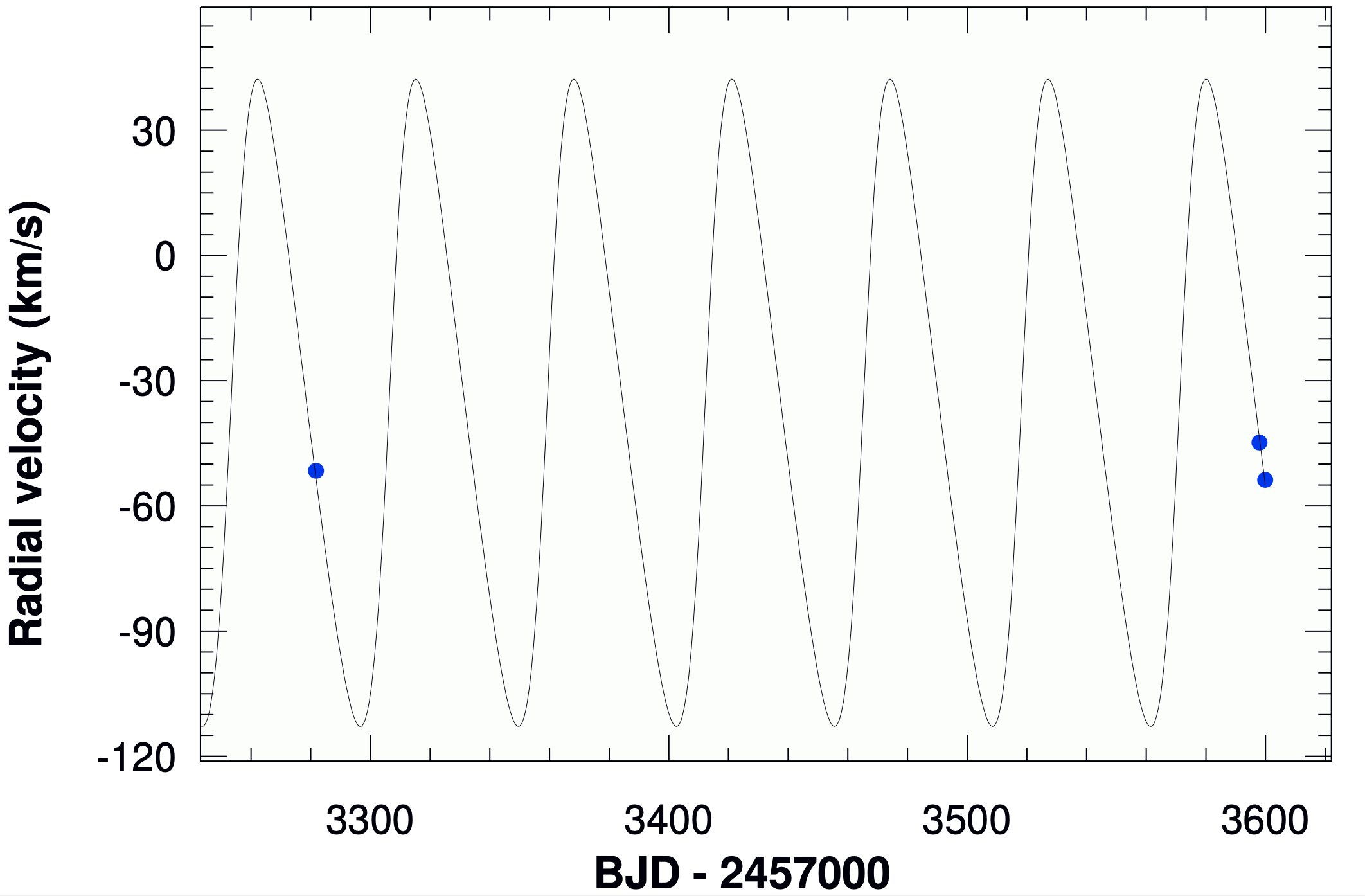}
\caption{
 {\it Left:} RV values reported by \citet{Eisner2022} fitted with $P_{\rm orb}=52.976\pm0.022$\,d.
{\it Right:} Our RV values obtained using PEPSI spectra fitted with the same period.
The measurements are denoted by blue dots and the fit for $P_{\rm orb}=52.976\pm0.022$\,d is highlighted by a black solid line.
}
   \label{fig:per}
\end{figure*}

\begin{figure*}
 \centering
\includegraphics[width=0.30\textwidth]{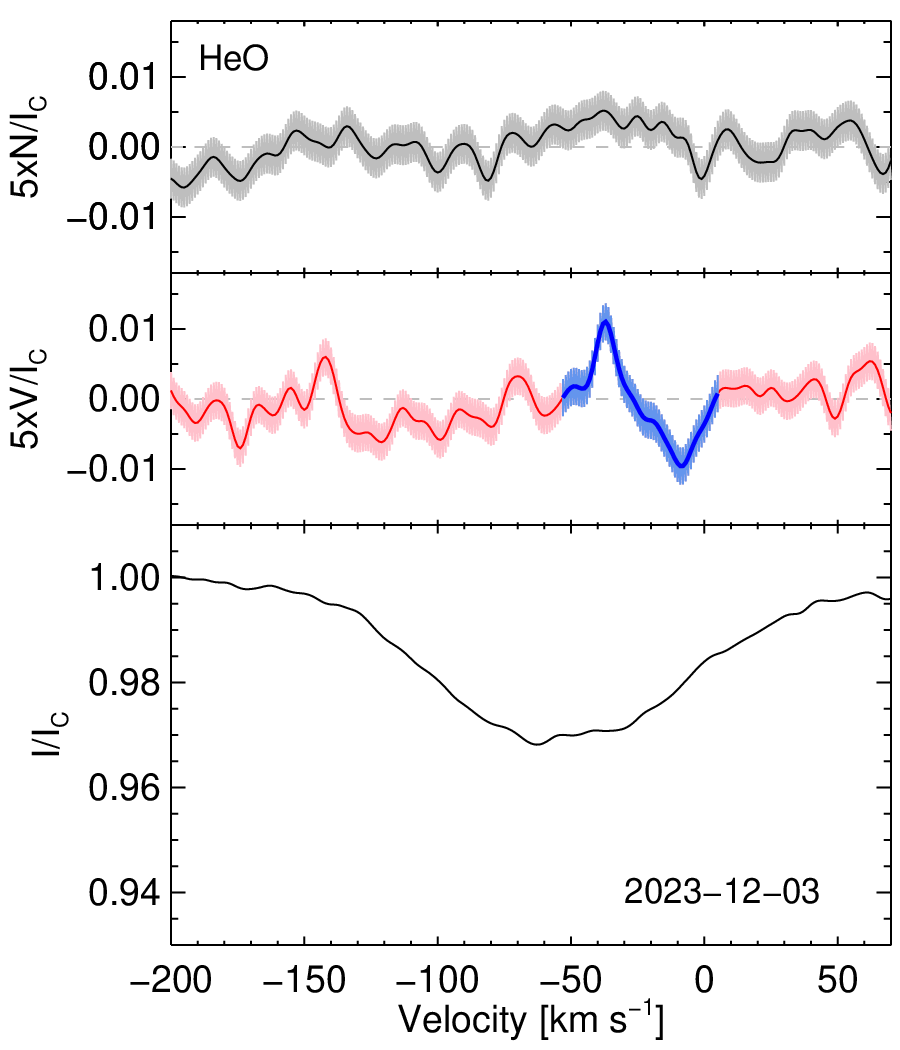}
\includegraphics[width=0.30\textwidth]{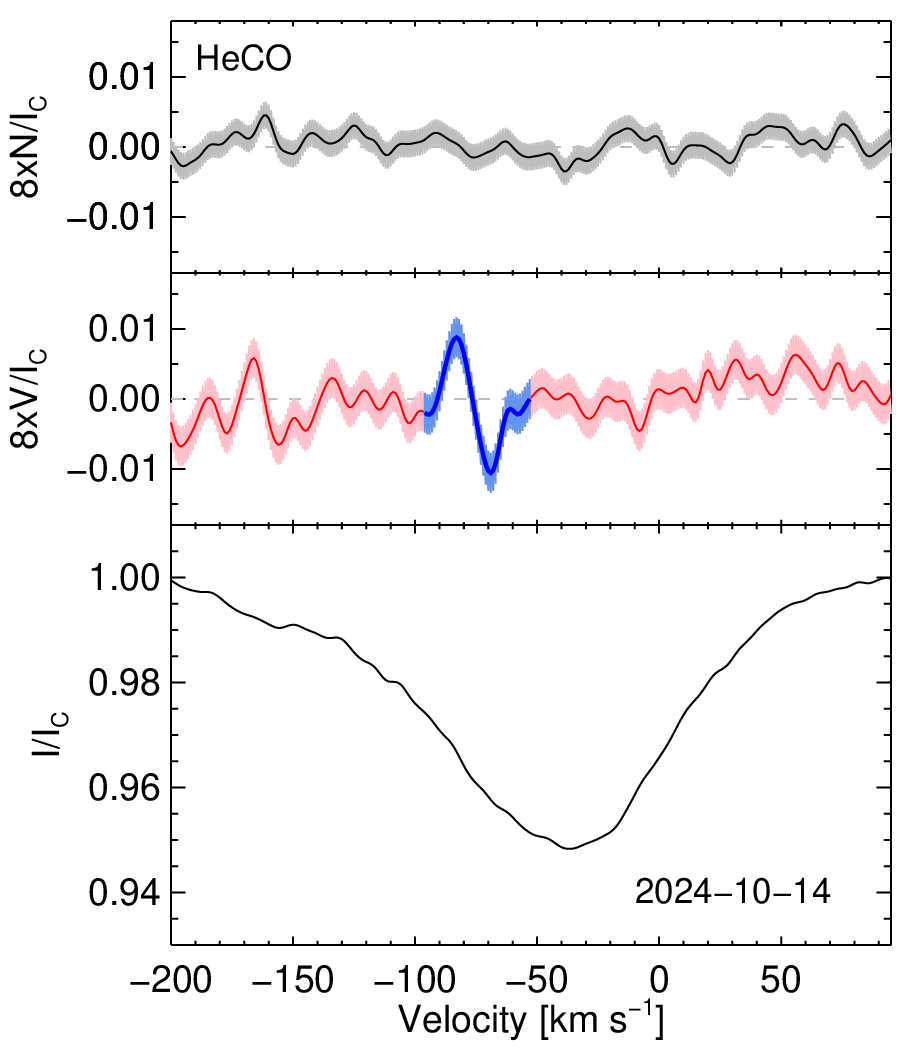}
\includegraphics[width=0.30\textwidth]{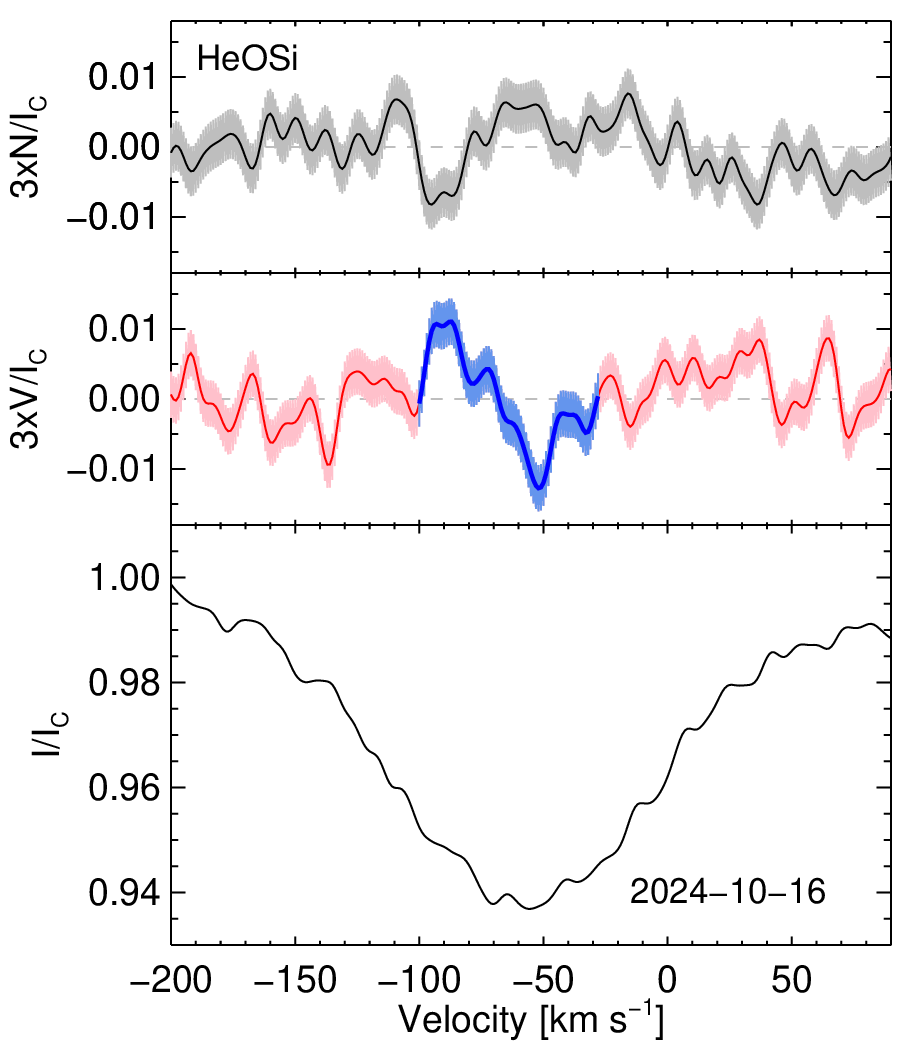}
\caption{
LSD Stokes~$I$, $V$, and diagnostic null $N$ spectra (from bottom to top)
calculated for TIC\,470710327. Recognizable features identified in the Stokes~$V$ spectra
are highlighted in blue. The grey bands in the $N$ spectra  and red bands in the Stokes~$V$
spectra correspond to 1$\sigma$ uncertainties. 
}
   \label{fig:LSD}
\end{figure*}

PEPSI and its spectropolarimetric capabilities have been described in detail by \citet{Strassmeier2015}.
The observations in polarimetric mode with a spectral resolution of $R \approx 130,000$
have been obtained with the crossdispersers~II and IV on three epochs, on 2023 December 3, and on
2024 October 14 and 16, and cover the wavelength regions
4222--4769\,\AA{} and  5368--6283\,\AA{}. To obtain circularly polarised spectra, we used for each observation
two sub-exposures, one recorded at a retarder angle of 45\degr{} and  another one at 135\degr{}.
The data reduction was done using the software package 
``Spectroscopic Data Systems for PEPSI'' based on \citet{Ilyin2000}
and described in detail by \citet{Strassmeier2018}.

To measure the mean longitudinal magnetic field and to increase
the signal-to-noise ratio, we employed the least-squares deconvolution (LSD) technique.
The details of this technique, 
as well as how the LSD Stokes~$I$, Stokes~$V$, and diagnostic null spectra are calculated, were 
presented by \citet{Donati1997} and in our previous papers (e.g.\ \citealt{Hubrig2018,Jarvinen2020}).
The normalization of the spectra to the continuum level is identical to the procedure
described in detail by \citet{Hubrig2013} for HARPS.

To evaluate whether the detected features are spurious or definite detections, 
we followed the generally adopted procedure to use the false alarm probability (${\rm FAP}$) based on 
reduced $\chi^{2}$ test statistics \citep{Donati1992}:
the presence of a Zeeman signature
is considered as a definite detection (DD)
if ${\rm FAP} \leq 10^{-5}$,
as a marginal detection (MD) if  $10^{-5}<{\rm FAP}\leq 10^{-3}$,
and as a non-detection (ND) if ${\rm FAP}>10^{-3}$.

The results of our analysis  are presented in Table~\ref{tab:obsall},
where we indicate the BJD values at the middle of the exposures, $S/N$ values measured
in the Stokes~$I$ spectra in the spectral region around 6100\,\AA{}, the line list 
applied with the LSD technique,  the radial velocities (RVs) measured
using the centre of gravity method applied to the LSD profiles and the FAP values with the detection flags
DD, MD, or ND. In the two last columns of this table we display the results of our measurements of the
mean longitudinal magnetic field and the corresponding orbital phase calculated 
using the initial epoch ${\rm BJD}=2458880.39$ from \citet{Eisner2022}.
 As the addition of our three RV measurements using the PEPSI spectra extends the time base of all available RV measurements
to more than 1700\,d, we analysed all old and new RVs to improve the orbital period using the method described by \citet{Ilyin2000}. If we
use only the old RV measurements given in the table A1 of
\citet{Eisner2022} to fit a Keplerian orbit, we obtain $P_{\rm orb}=52.574\pm0.303$\,d, which is in agreement with the
Keplerian orbital period given by \citet[][cf. Fig.~6]{Eisner2022}, without taking into account their eclipse-timing variations.
Adding the new three RV data points measured using the PEPSI spectra changes the period to
$P_{\rm orb}=52.976\pm0.022$\,d.
The best fitting model for the RVs from \citet{Eisner2022} and for the RVs measured
using the PEPSI spectra is presented in Fig.~\ref{fig:per}.

Our LSD Stokes~$I$, $V$, and diagnostic null $N$ spectra calculated for all three PEPSI observations are presented in Fig.~\ref{fig:LSD}.
We detect definite magnetic fields on two epochs and a marginal field on the last epoch on 2024 October 16.
All measured field strengths show positive polarity.
Assuming a dipolar magnetic field configuration -- which is typical for many massive stars
with observed magnetic fields --
and a limb-darkening factor of 0.3 \citep{Claret2011}, we can estimate the minimum of the dipole strength
$B_{\rm d}\ge3.53\left< B \right>_{\rm z,max}$ following \citet{Preston1967}.
With $\left< B \right>_{\rm z,max}=709$\,G we obtain $B_{\rm d}\ge2.5$\,kG.

In agreement with \citet{Eisner2022}, only spectral lines belonging to the massive tertiary are identified in our
observations, suggesting that the magnetic field is detected in this component.
However, the presence of the close companion TIC\,470710327$^\prime$ at an angular separation of
$\sim0.5$\,arcsec and with $V=11.6$ has been reported
in a number of previous speckle interferometric observations and was also confirmed by the more recent
observations by \cite{Eisner2022}, shown in their Figure~1.
That figure, however, illustrates two problems with the
phase reconstruction of the image, a second object at the opposite site of the central
star (a residual from the autocorrelation function), and two wide (in the image black) crosses centred on the
companion and its counterpart.
Both issues will have an impact on estimating the intensity
ratio between TIC\,470710327 and TIC\,470710327$^\prime$, if this
was indeed performed on either the image or on its representation
in Fourier space.
It is quite possible that the error in the intensity
ratio introduced by these two features is larger than 10\% and
that this needs to be taken into account in future discussions of this system.

Further, \citet{Eisner2022} made use of more than 4000 photometric measurements of
TIC\,470710327 obtained with ground based telescopes.
These observations are very likely all seeing limited and should
have a spatial resolution around or above 1\,arcsecond, independent
of the actual pixel scale used.
Thus, TIC\,470710327$^\prime$ is not excluded in these data,
unlike stated by \citet{Eisner2022}.
There is also a discrepancy between the parallax of $4-4.5$\,kpc determined
by \citet{Eisner2022} and the parallax of $\sim 1$\,kpc from Gaia
that has to be readdressed once a new Gaia data release becomes available
that addresses astrometric signals from binaries and higher multiples.

Since for the PEPSI observations the aperture on the sky of $\sim1.5$\,arcsec includes the component
TIC\,470710327$^\prime$, it
is possible that the light of this component contaminates, and thus dilutes, the spectrum of the tertiary.
In case of contamination by TIC\,470710327$^\prime$ in the PEPSI observations,
we would expect that the amplitudes of the observed Zeeman features are lower
in comparison to the size of these features in the spectra without light contamination
(e.g.\ \citealt{Hubrig2023}). As a consequence, the magnetic field of the tertiary is expected to
be even stronger than the estimated lower boundary of the magnetic field of 2.5\,kG.

\section{Discussion}
\label{sec:disc}

Recently, Sim\'on-D\'iaz (priv.\ comm.) carried out a line-broadening analysis with the IACOB-BROAD software tool
\citep{Simon2014} using the \ion{Si}{III}~4567 line identified in the spectra of the tertiary. The Fourier transform
(the FT method) and the goodness-of-fit (the GOF method)
analysis resulted in a quite low projected rotation velocity $v \sin i\approx56$\,km\,s$^{-1}$ and a macroturbulent
broadening $v_{mac}$ of about
80\,km\,s$^{-1}$. The FT method directly estimates $v \sin i$ from the position of the
first zero in Fourier space, whereas the GOF method convolves
synthetic line profiles for a range of projected rotational and macroturbulent velocities, creating a
standard  $\chi^{2}$--landscape from which a best combination of the two parameters is determined
\citep{Simon2014}. 
It is intriguing that the measured $v_{mac}$ is quite large and typical for massive supergiants
(e.g.\ \citealt{Simon2010, Simon2017}). Interestingly, it is  also typical for magnetic O-type stars (see table F4 in
\citealt{Holgado2022}).
According to \citet{Simon2017}, to explain the presence of additional line broadening, 
heat-driven non-radial modes and spectroscopic variability caused by rotational modulation, strange modes, stochastically
excited non-radial modes and/or convectively driven internal gravity waves can be considered.

Also, recent studies of massive stars with magnetic fields showed that they rotate much slower than their non-magnetic
counterparts. The longest rotation period was reported for the O8f?p star HD\,108 with a period of about 54\,yr
(e.g.\ \citealt{Naze2001}). 
Rotation periods of magnetic stars are usually determined using the rotational modulation of the measured mean longitudinal
magnetic field with magnetic phase curves showing a sinusoidal character for dipole field configurations, or
using rotational modulation of emission observed in hydrogen lines, or spectral line intensities (e.g.\ \citealt{Grunhut2012}).

The rotation period of the tertiary in the system TIC\,470710327 is currently unknown.
On the other hand, because the spectral classification
O9.5V-B0.5V and the corresponding mass range of 14.5--16\,$M_{\odot}$ were already determined by \citet{Eisner2022}, we
can estimate the radius of the tertiary making use of
the tabulation of \citet{Harmanec1988} based on the compilation of data on masses and radii of stars. With an estimated
radius of 5.6$R_{\odot}$, a $v \sin i\approx56$\,km\,s$^{-1}$, and an inclination of about $67\degr$ (assuming that the
axis of rotation is identical with the orbital inclination of the wide orbit), the rotation period should be about 5.5\,d. Using
this period our second and third PEPSI observations respectively correspond to the phases 0.49 and 0.85,
considering the date of the first observations as initial epoch.

\begin{figure*}
 \centering
\includegraphics[width=\textwidth]{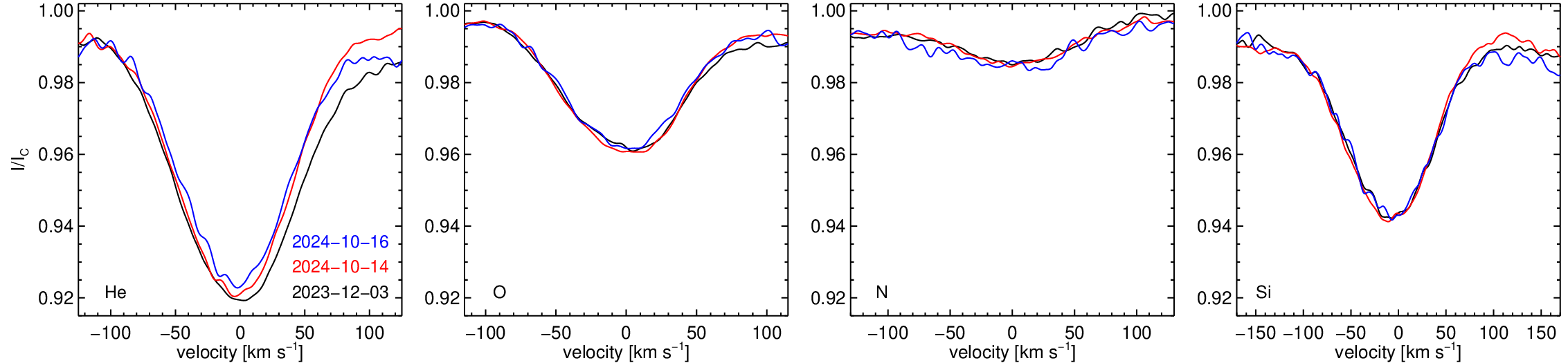}
\caption{
The LSD line profiles belonging to different elements observed in the PEPSI spectra of TIC\,470710327 on three
different epochs.
Apart from minor changes in the He line profiles, no significant changes in other line profiles are detectable.
The weakness of the nitrogen lines is notable.
         }
   \label{fig:var}
\end{figure*}

To check the spectral variability of the tertiary on three different observing
epochs, we present in Fig.~\ref{fig:var} the LSD line profiles belonging
to different elements. Unfortunately, H$\alpha$ and H$\beta$ are not recorded in the PEPSI setup used for our observations.
We do not detect significant changes of the line profiles apart from minor
changes in the line profiles of the He lines.
Noteworthy, in contrast to widespread expectations of nitrogen enrichment in magnetic massive stars,
we show in Fig.~\ref{fig:var} that the nitrogen lines in the spectra of TIC\,470710327 are extremely weak.

To our knowledge, two other less massive triple systems with a magnetic tertiary have previously been considered in the literature.
Over a decade ago, using the ESO FORS2 and HARPS instruments, \citet{Hubrig2013} detected a strong magnetic field of about 600\,G in the poorly
studied multiple system HD\,164492C located in the Triﬁd nebula, one of the youngest star
forming regions. Later on, \citet{Gonzalez2017} showed that the system is formed by an eccentric close spectroscopic binary
with a period of 12.5\,d, and a massive tertiary with a separation of the order of 100\,au. 
The tertiary component possesses a strongly variable magnetic field with a most
probable period in the range of 1.4--1.6\,d. With an effective temperature of about 25,000\,K and a mass of about 11\,$M_{\odot}$,
it is apparently the fastest rotator and the most massive star of this triple system and has anomalous chemical
abundances with a marked helium overabundance of $0.35\pm0.04$.

V746\,Cas is known to be a triple system with a B-type primary in a close binary with a period of 25.4\,d and
a distant cooler B-type third component in a 170\,yr (62,000\,d) orbit \citep{Harmanec2018}.
The magnetic signatures corresponding to the longitudinal magnetic field varying between about –200 and 200\,G
seemed to be associated with the primary component \citep{Neiner2014}. In contrast, \citet{Harmanec2018} showed
that the distant tertiary component is magnetic and is moving in a long orbit with the 25.4\,d binary. Moreover, these authors
showed that the magnetic field of the tertiary varies with a photometric period of 2.504\,d, which could almost certainly 
be identified with its rotational period.

Two massive targets have been most frequently discussed in the literature as merger products based on
the observational evidence.
\citet{NievaPrzybilla2014} measured the stellar parameters of the magnetic massive B0.2~V star $\tau$~Sco belonging to the
Upper Sco association and, based on its position in the H-R diagram, concluded that it is a blue straggler star
and could possibly originate from a stellar merger. 
\citet{Frost2024} combined ten years of spectroscopic and interferometric data for the
magnetic Of?p binary
HD\,148937 with an orbital period of 29\,yr and reported that the magnetic primary, although more massive, appears
younger, suggesting that a merge or mass transfer took place in this system during earlier evolution.

In our study we build upon the conjecture presented by \citet{Vigna2022}, who proposed that the most massive star in
TIC\,470710327 is a merger product. Our analysis confirms that this star is magnetic and slowly rotating, as predicted. 
However, we did not detect any nitrogen enhancement or other abnormal abundances, which would further support the
hypothesis for the merger scenario. \citet{Vigna2022} suggested that the merger might have occurred dynamically
through von Zeipel-Lidov-Kozai oscillations \citep{vonZeipel1910,Lidov1962,Kozai1962},
a scenario that could be further investigated if the individual stellar spins of TIC\,470710327 are eventually determined. 
Alternatively, the merger could have taken place late on the pre-main sequence or shortly after the zero-age main sequence,
which might explain the absence of peculiar abundances in the merger product.
The reduction of TIC\,470710327 from a quadruple to a triple system may be a common evolutionary pathway
for high-mass stars  \citep[e.g.][]{Preece2024}. 
Further characterisation of TIC\,470710327 and future observations of compact multiple-star systems could shed light
on both the primordial multiplicity of massive stars and the properties of merger remnants.

\section*{Acknowledgements}

We thank the referee Peter Harmanec for very useful suggestions. 
We also thank Sergio Sim\'on D\'iaz, who estimated
for TIC\,470710327 the $v \sin i$-value and the macroturbulent broadening $v_{\rm mac}$.
This work is based on observations carried out with the PEPSI spectropolarimeter.
PEPSI was made possible by funding through the State
of Brandenburg (MWFK) and the German Federal Ministry of Education and
Research  (BMBF)  through  their  Verbundforschung  grants  05AL2BA1/3  and
05A08BAC. LBT Corporation partners are
the University of Arizona on behalf of the Arizona university system;
Istituto Nazionale di  Astrofisica,  Italy;
 LBT  Beteiligungsgesellschaft,  Germany,  representing  the
Max-Planck Society, the Leibniz-Institute for Astrophysics Potsdam (AIP), and
Heidelberg  University; the  Ohio  State  University; and  the  Research  Corporation,
on behalf of the University of Notre Dame, the University of Minnesota and
the University of Virginia. 

\section*{Data Availability}

The PEPSI data can be obtained from the authors upon reasonable request.

   \bibliographystyle{mnras} 
   \bibliography{tertiary4} 

\bsp	
\label{lastpage}
\end{document}